\documentclass[reprint,amsmath,amssymb,aps,pre,superscriptaddress,a4paper]{revtex4-2}
\usepackage{graphicx}
\usepackage[colorlinks=true,allcolors = blue]{hyperref}
\usepackage{newtxtext,newtxmath}
\usepackage{microtype}
\usepackage{selnolig}
\usepackage{siunitx}
\newcommand{\order}[1]{\mathcal{O}({#1})}

\newcommand{\dd}{\mathrm{d}}

\newcommand{\smallfrac}[2]{{\textstyle\frac{#1}{#2}}}

\renewcommand{\Re}{\mathrm{Re}} 
\newcommand{\St}{\mathrm{St}} 
\newcommand{\Stcrit}{\St_{\mathrm{c}}}

\usepackage{bm}
\renewcommand{\vec}[1]{\bm{\mathbf{#1}}}

\newcommand{\Eq}[1]{Eq.~\eqref{#1}}
\newcommand{\Eqs}[1]{Eqs.~\eqref{#1}}
\newcommand{\Fig}[1]{Fig.~\ref{#1}}
\newcommand{\partfig}[2]{\hyperref[#1]{#2}}
\newcommand{\partFig}[2]{Fig.~\hyperref[#1]{\ref*{#1}#2}}

\newcommand{\Refcite}[1]{Ref.~\cite{#1}} 

\newcommand{\naive}{{na\"\i{}ve} }

\newcommand{\latin}[1]{{\itshape #1}}
\newcommand{\etal}{\latin{et al.}}

\usepackage{todonotes}
\setuptodonotes{inline, color=blue!15}

\usepackage{xr-hyper}
\graphicspath{{figures/}}

\begin{document}

\title{Critical inertia for particle capture is determined by surface geometry at forward stagnation point}

\date{\today}

\author{Joshua F. Robinson}
\email{joshua.robinson@stfc.ac.uk}
\affiliation{STFC Hartree Centre, Sci-Tech Daresbury, Warrington, WA4 4AD, United Kingdom}
\affiliation{H.\ H.\ Wills Physics Laboratory, University of Bristol, Bristol BS8 1TL, United Kingdom}
\affiliation{Institut f\"ur Physik, Johannes Gutenberg-Universit\"at Mainz, Staudingerweg 7-9, 55128 Mainz, Germany}

\author{Patrick B. Warren}
\email{patrick.warren@stfc.ac.uk}
\affiliation{STFC Hartree Centre, Sci-Tech Daresbury, Warrington, WA4 4AD, United Kingdom}
\affiliation{SUPA and School of Physics and Astronomy, The University of Edinburgh, Peter Guthrie Tait Road, Edinburgh EH9 3FD, United Kingdom}

\author{Matthew R. Turner}
\email{m.turner@surrey.ac.uk}
\affiliation{School of Mathematics and Physics, University of Surrey, Guildford, GU2 7XH, United Kingdom}

\author{Richard P. Sear}
\email{r.sear@surrey.ac.uk}
\homepage{https://richardsear.me/}
\affiliation{School of Mathematics and Physics, University of Surrey, Guildford, GU2 7XH, United Kingdom}

\begin{abstract}
  Aerosols are ubiquitous, and particle capture from particle-laden air as it flows past an obstacle is of widespread practical importance.  Neglecting diffusion, previous work has shown that for a smooth curved surface in both Stokes flow and inviscid flow, only particles with inertia above a threshold value (quantified by the nondimensional Stokes number) collide with the surface.  Here we show that the critical Stokes number decreases with increasing Reynolds number of the air flow, and the mechanism behind this threshold is the same at all finite Reynolds numbers but becomes qualitatively different in the limit of infinite Reynolds number (inviscid flow).  In addition we show that in the latter case (inviscid flow) the threshold is set solely by the flow near the stagnation point, whereas at finite Reynolds numbers the threshold also depends on the flow far from the stagnation point. The threshold also depends on obstacle geometry and we show that fibers whose cross section is flattened along the flow direction have greater size selectivity than fibers with a circular cross section.
\end{abstract}


\maketitle

\section{Introduction}

The air we breath is an aerosol; it contains suspended particles, with sizes up to tens of micrometres. When air flows around an obstacle, aerosol particles may collide with and so deposit on, the obstacle. This can cause problems, such as freezing water droplets causing the icing of aircraft wings and wind turbine blades \cite{taylor1940,*taylor_vol3,langmuir1946,parent2011}. It can also be useful, face masks and other air filters rely on aerosol particles depositing on their (typically $\sim \SI{10}{\micro\metre}$) fibers \cite{wang2013,howard2020,greenhalgh2020,riosdeanda2022,robinson2022,robinson2021} to remove these aerosol particles from the air. Other applications include aerosolised drugs \cite{finlay2019,*darquenne2020,*chalvatzaki2020,*cheng2014,*zhang2020}, the wind-pollination of plants (anemophily) \cite{pawu1989,*niklas1985}, and the harvesting of water from fogs by plants and animals \cite{parker2001,*mitchell2020,*shahrokhian2020}, and by humans \cite{azeem2020}. There are also industrial \cite{zamani2009} and geological applications \cite{vasconcelos2009}.
The same physics also occurs when particles suspended in a flowing liquid, such as sea water, deposit on an obstacle, for example, as happens with coral feeding on plankton \cite{boudina2020,*espinosa-gayosso2021,*palmer2004}. It is worth noting that these varied flows cover the complete range of Reynolds numbers. In air filters it is much less than one, i.e.\ we have  Stokes flow.  With aircraft wings and wind turbine blades though, the Reynolds number is typically much larger (not least because the length scale is enormously bigger).

Here we study collisions of a particle with an obstacle, or `collector', when the collisions are driven by the inertia of the aerosol particle, which causes it to deviate from the streamlines of the air flow. These streamlines flow around the obstacle so a point particle that follows streamlines perfectly always goes around an obstacle having without colliding with it. Particles with inertia do not follow the streamlines, meaning they can collide with an obstacle's surface. In 1931, \textcite{albrecht1931} found a threshold in the inertia, below which no (point) particles deposited on the obstacle. A minimum amount of inertia is needed before any (point) particles are deposited. Then in the 1940s, first \textcite{taylor1940}, and then Langmuir and Blodgett \cite{langmuir1946} calculated this critical value of the inertia. The particle inertia is characterised by the Stokes number, $\St$, and the threshold is at some critical value $\St=\Stcrit$.

There has been considerable work on this problem since the works of Taylor \cite{taylor1940,*taylor_vol3}, Langmuir and Blodgett \cite{langmuir1946} in the 1940s \cite{finstad1988comp,finstad1988median}, motivated by its many applications. 
How the threshold varies with Reynolds number has been studied by Phillips and Kaye \cite{phillips1999}, and Ara\'{u}jo~\etal~\cite{araujo2006} computed the deposition efficiency on cylinders in low Reynolds number flow. They found
that the critical scaling for the deposition efficiency was $\lambda\propto (\St-\Stcrit)^{1/2}$.  Vall\'{e}e~\etal~\cite{vallee2018} studied deposition on spheres, using analytics and computationally. They determined both the critical value of the Stokes number as a function of the Reynolds number of the flowing air, and the critical scaling of the deposition efficiency. At low Reynolds number they showed that the critical scaling is $\lambda\propto (\St-\Stcrit)^1$. Deposition occurs on a line for Ara\'{u}jo~\etal's cylinders and on an area for Vall\'{e}e~\etal's spheres, so the exponent found by Vall\'{e}e~\etal~\cite{vallee2018} should be, and is, twice that found by Ara\'{u}jo~\etal~\cite{araujo2006}.  As Vall\'{e}e~\etal~\cite{vallee2018} first showed, the critical scaling is very different for inviscid (high Reynolds number) flow around a sphere. The scaling is not a power law but the unusual behaviour $\lambda\propto \exp(-1/(\St-\Stcrit)^{1/2})$. This also applies for a cylinder \cite{turner2023}.

Here we characterise the dynamical transition that sets the critical deposition threshold at $\St=\Stcrit$, and show how it is determined by the flow field near the forward stagnation point at the front of the obstacle. This stagnation point (e.g.\ the white circle in \partFig{fig:kuwabara}{a}) is a point where the flow velocity is zero and where the flow divides to go around the obstacle. We also show why, just above the threshold, the amount of aerosol deposition scales as $(\St-\Stcrit)^{1/2}$, for air flow at finite Reynolds numbers, Re, but not for inviscid ($\Re \to \infty$) flow. It is important to understand aerosol deposition as a function of Reynolds number as in applications Re varies from less than one (masks) to millions (aircraft wings).

\begin{figure}
  \centering
  \includegraphics[width=\linewidth]{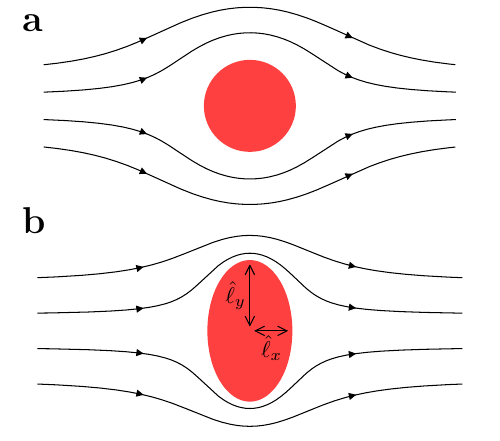}
  \caption{
    (colour online)
    Fluid flows around cylindrical collectors with (a) circular and (b) elliptical cross-sections.
    These collectors represent idealised obstacles in filters.
  }
  \label{fig:ellipse}
\end{figure}

\begin{figure}[b]
  \centering
  \includegraphics[width=\linewidth]{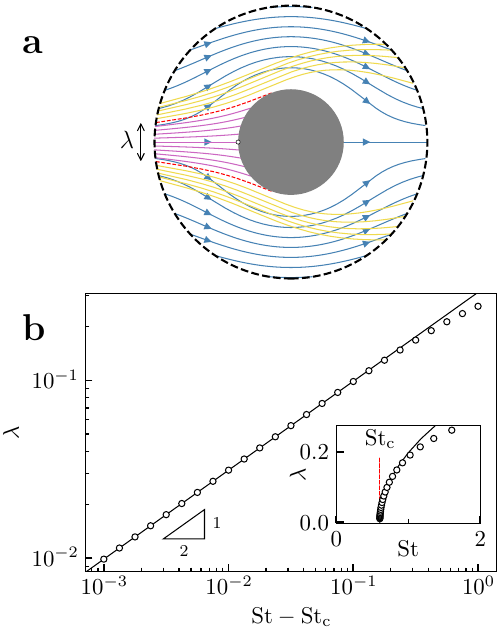}
  \caption{
    (colour online)
    Particle capture by a cylindrical collector at zero Reynolds number flow (via the Kuwabara flow field with volume fraction $\alpha=0.15$).
    (a) Representative trajectories at $\St=2$ in the Kuwabara flow field (light blue) which hit (purple) or miss (yellow) the cylindrical collector (grey).  Trajectories are launched with zero initial acceleration from the outer circular perimeter of the Kuwabara flow field where the vorticity vanishes (dashed).
    The forward stagnation point is shown with a white circle.
    (b) Capture efficiency $\lambda$ as a function of Stokes number $\St$\,; the points are from numerical experiments and the solid line is $\lambda\propto\sqrt{\St-\Stcrit}$ with $\Stcrit\simeq0.604$.
  }
  \label{fig:kuwabara}
\end{figure}

Defining a Stokes number requires a length scale from the (air flow round the) obstacle. For a long cylinder normal to the flow there is only one length scale, the radius $R$, see \partFig{fig:ellipse}{a}, and this sets the local curvature at the stagnation point. To better understand how to control the critical Stokes number and collection efficiency of an obstacle, we generalise from a circular cross-section to an elliptical cross-section, see \partFig{fig:ellipse}{b}.

Ellipses, like discs, are smooth curved bodies, and there are well known analytic expressions for inviscid flow around ellipses using conformal mapping techniques \cite{acheson_book}. Unlike a disc, which has just one length scale, ellipses have two. These can be of a few different pairs, one example pair is the half-length along the flow direction, $\hat{\ell}_x$, and half-length normal to that direction, $\hat{\ell}_y$ as in \partFig{fig:ellipse}{b}. Here we only consider ellipses with a symmetry axis (semi-major or semi-minor) along the flow direction.

Most of the work for high Reynolds numbers has been computational or theoretical, as experiments on collisions in these flows are challenging. However,
Wong and coworkers did obtain some experimental data on deposition efficiency at Reynolds numbers of a few hundreds \cite{wong1955}. This was for an aerosol with particles with a narrow distribution of sizes. They found no measurable deposition below a threshold near that predicted by Taylor \cite{taylor1940,*taylor_vol3}, and by Langmuir and Blodgett \cite{langmuir1946}. So these experiments agree with theoretical predictions (for a simplified model) that a threshold exists. Makkonen and coworkers \cite{makkonen1987,*makkonen1992,*makkonen2018} have measured ice deposition on cylinders. This is for the typical case in the environment, where the droplets have a broad range of sizes, which complicates comparison with theoretical predictions.

In section \ref{sec:model} we present our model for inertial particles impacting on elliptical obstacles.
We then analyse how critical inertia emerges in two distinct limits in section~\ref{sec:on-axis}.
In section~\ref{sec:off-axis} we examine how the efficiency of particle capture varies above the critical threshold in each limit, and how this is affected by elliptical eccentricity.
Last, in section~\ref{sec:rg} we discuss some recent work where the critical inertia can be calculated analytically, before concluding with some perspectives on filters.
All calculations underlying this work are available in \Refcite{rimeflows}.

\section{Model}
\label{sec:model}

Here we consider a simple model: point particles that interact with the obstacle only when they collide. Real particles have a size and exhibit more realistic interactions with the obstacle (e.g.\ electrostatics in masks made to N95/FFP2 standards \cite{wang2013}). However, aerosol particles are typically no more than tens of micrometres in size and in some cases the obstacle can be tens of centimetres across, so the particles are much smaller than the obstacle and thinking of them as point particles is a good approximation. For air filters (including masks) the obstacle diameter is much smaller and can be comparable to the particle size. Here particle size does affect deposition \cite{wang2013,robinson2021,riosdeanda2022} but the dynamical transitions we study here may well underlie some counterintuitive behaviour seen for particles with a non-zero diameter: Under some circumstances, increasing particle inertia {\em decreases} the amount of deposition \cite{fernandezdelamora1982,robinson2021,riosdeanda2022}.

We study a simple model of point particles depositing onto bodies that are infinite along the $z$ axis and either circular or elliptical in the $x-y$ plane. So our problem is two dimensional. The ellipses are orientated perpendicular to the flow as shown in \partFig{fig:ellipse}{b}.

As Taylor \cite{taylor1940,taylor_vol3}, and Langmuir and Blodgett \cite{langmuir1946} realised, particle capture occurs within a stagnation point flow.
Stagnation points are loci where the flow velocity is zero.
Every point along a surface where a no-slip boundary condition is applied in viscous fluid flow could be considered as a stagnation point, but for our purposes we only use ``stagnation point'' to refer to such points where the flow speed would be zero regardless of boundary condition.
Ellipses have stagnation points at the mid-point of their surface where the flow is bisected. 
See the horizontal streamlines in \partFig{fig:kuwabara}{a} for the case of a cylinder.

For a particle suspended in flowing air we assume that the only force on the particle is the friction with the surrounding air, which is taken to be proportional to the difference between particle's velocity $\hat{\vec{v}}$ and the local flow velocity $\hat{\vec{u}}$. This force is taken to act on the particle's centre of mass. Then
Newton's equation for the particle motion is
\begin{equation}\label{eq:stokes-newton}
  m \frac{\dd \hat{\vec{v}}}{\dd \hat{t}} = - \xi(\hat{\vec{v}} - \hat{\vec{u}})
\end{equation}
for a particle of mass $m$ and with drag coefficient $\xi$. We use a hat (\,$\hat{}$\,) to indicate a dimensional quantity.  We shall also need a characteristic flow speed $U$, which is discussed further below.
We switch to nondimensional quantities via transformations $\vec{r} = \hat{\vec{r}} / \hat{\ell}_y$, $\vec{u} = \hat{\vec{u}} / U$ and $t = U \hat{t} / \hat{\ell}_y$ where $2 \hat{\ell}_y$ is the cross-sectional width of the ellipse in the plane transverse to the incoming flow, see \Fig{fig:ellipse}.  This leads to
\begin{equation}\label{eq:stokes-newton-nondim}
  \St \, \dot{\vec{v}} + \vec{v} - \vec{u}
  \equiv
  \St \, \ddot{\vec{r}} + \dot{\vec{r}} - \vec{u} = 0\,,
\end{equation}
where $\St = m U / (\hat\ell_y \xi)$ is the Stokes number giving the effective inertia, and the dot (\,$\dot{}$\,) indicates derivative with respect to nondimensional time $t$.
In our dimensionless units we have effective ellipse dimensions $\ell_x = \hat\ell_x / \hat\ell_y$ and $\ell_y = \hat\ell_y / \hat\ell_y \equiv 1$.
The circular case is recovered by setting $\ell_x = 1$ in these units. 

We are interested in particle trajectories that pass close to the forward stagnation point.
We take the origin to be at the centre of the obstacle so that this stagnation point occurs at $(r, \theta) = (\ell_x, \pi)$ in polar coordinates.
In polar coordinates we must include inertial terms in the acceleration term of
\Eq{eq:stokes-newton-nondim}.
Decomposing $\vec{u} = u_r\vec{e}_r + u_\theta \vec{e}_\theta$ where $\vec{e}_r$ and $\vec{e}_\theta$ are unit vectors in the radial and azimuthal directions respectively, we find
\begin{equation}\label{eq:stokes-newton2}
  \begin{split}
    \St(\ddot{r} - r\dot\theta^2)
    &= - (\dot{r} - u_r)\,,\\
  \St(r\ddot{\theta} + 2\dot{r}\dot{\theta})
  &= - (r\dot{\theta} - u_\theta)\,.
  \end{split}
\end{equation}

For the the original \Eqs{eq:stokes-newton} there is often a natural choice of $U$ for any given problem, e.g.\ the flow speed infinitely far from a cylinder $U_\infty$.
Fernandez de la Mora and Rosner \cite{fernandezdelamora1982} have pointed out that alternative choices of $U$ can be made, giving different Stokes numbers for the same system.
The particular choice selects a distinct different asymptotic regime of validity \footnote{This situation is analogous to how the Reynolds number varies with length scale in classical problems considering the flow past an object (e.g.\ sphere or cylinder). These singular problems prompted the development of asymptotic techniques. See e.g.\ \Refcite{veysey2007} for discussion and historical context.}.
Following these authors \cite{fernandezdelamora1982}, we take our characteristic flow speed to be the magnitude of the \emph{local} flow approaching the stagnation point.
With this choice the normal component of flow expands about the stagnation point as
\begin{equation*}
  u_r = - (r-\ell_x)^m + o\Big( \big(r - \ell_x\big)^m \Big)
\end{equation*}
as $r \to \ell_x$, with exponent $m > 0$ ensuring no penetration.
Note in the case of inviscid flow around the circle/ellipse this choice involves taking $U = 2 U_\infty$, so our Stokes number will be double that of authors taking $U = U_\infty$.

\section{On-axis flows and the critical Stokes number}
\label{sec:on-axis}

As Taylor \cite{taylor1940,taylor_vol3}, and Langmuir and Blodgett \cite{langmuir1946} realised, there is a threshold inertia, and so Stokes number, below which no point particles collide with an obstacle. The particles all follow the fluid flow around the obstacle. This threshold or critical Stokes number, $\Stcrit$, depends only on the on-axis flow (i.e.\ with constant $\theta=\pi$) \cite{taylor1940,*taylor_vol3,langmuir1946,vallee2018}. In this section we will compute values for $\Stcrit$.

\subsection{On-axis flow fields and equations of particle motion}
\label{sec:on-axis-flows}

On-axis ($\theta=\pi$) the particle equation of motion reduces to a single differential equation for $r(t)$.
It is convenient to introduce the auxiliary variable $x = r - \ell_x \in [0, \infty)$ quantifying the distance from the forward stagnation point.
Below the critical Stokes number $\Stcrit$ there are no collisions in finite time.
The on-axis particle will eventually reach the surface at $x=0$ but only in the limit $t\to\infty$.
By contrast, the particle reaches $x = 0$ in finite time for $\St > \Stcrit$ which we call a collision.

On-axis, the equation of motion \Eq{eq:stokes-newton2} reduces to
\begin{equation}
  \St \, \ddot{x}=
  - \left( \dot{x} - u_r \right)\,.
  \label{eq:oned1}
\end{equation}
We consider three possible on-axis flow fields $u_r$
\begin{equation}
    u_r(x)
    =
    \begin{cases}
      -x^2  & \mbox{incompressible Stokes flow}  \\
     -x   & \mbox{inviscid flow}\\
     -x^m & \mbox{generalised flow}
     \end{cases}.
    \label{eq:xffs}
\end{equation}
Note that the proportionality constant is unity in each case from our choice of $U$ in the definition of $\St$ (see end of previous section).

The quadratic scaling of the first flow field $u_r \propto -x^2$ is generic to incompressible flows where a no-slip boundary condition is applied.
The specific case we will consider is the leading order term in a small $x$ expansion in the vicinity of the stagnation point for Stokes flow around a cylinder.
The so-called ``Stokes paradox''  \footnote{This misleadingly-named phenomenon is not truly a paradox. See \cite{veysey2007} for a comprehensive discussion and historical account.} prohibits Stokes flows around isolated cylinders \cite{vandyke1975}, so the underlying flow field must be imagined as a finite pore in a dense media.
As a specific flow field we take the Kuwabara flow which was developed to approximate an array of equally spaced cylinders \cite{kuwabara1959}.
Note that the constant of proportionality in this case depends (weakly) on the density of cylinders.
The Kuwabara flow field and its expansion is given in section 1 of the Supplemental Material (SM).

The second flow field is the leading order term for inviscid flow near a stagnation point on the surface of an ellipse.
More generally, a linear leading flow is expected wherever incompressibility or the no-slip boundary condition is relaxed.
The flow field for inviscid flow past an ellipse is given in section 3 of the SM.
The final flow field with $m\in[1,2]$ is a generalised near-stagnation-point flow field, which we use to continuously vary the limiting flow past a cylinder between the inviscid limit at $m=1$ and the Stokes limit at $m=2$.

\begin{figure*}
  \centering
  \includegraphics[width=\linewidth]{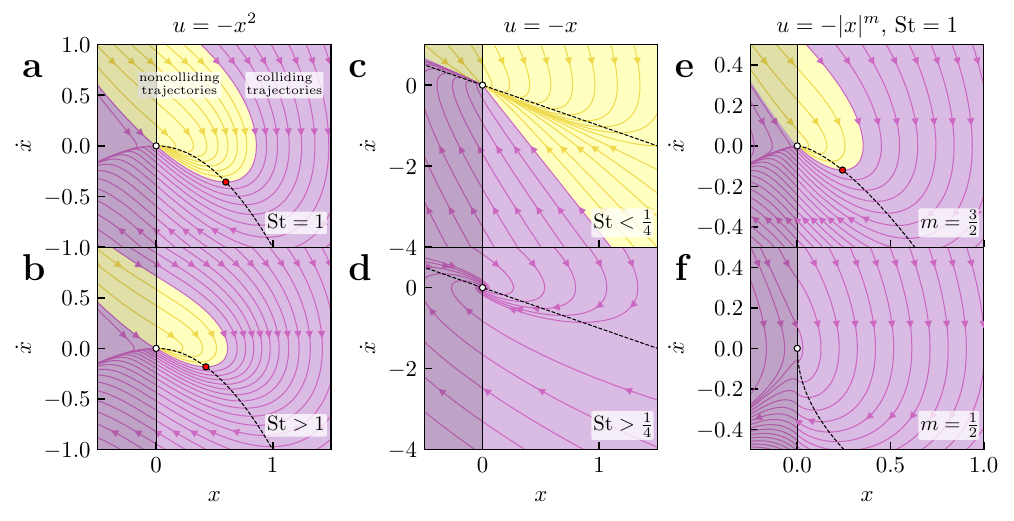}
  \caption{
  (colour online) Phase space portraits for the on-axis problem in the $(x, \dot{x})$-plane for (a-b) Stokes flow, (c-d) inviscid flow, and (e-f) more general power law flows (noninteger $m$). (a) to (b), and (c) to (d), the Stokes number is increased, while from (e) to (f) the exponent $m$ increases at fixed Stokes number. In all panels, trajectories that collide are shown in yellow, while trajectories that do not collide (except at $t\to\infty$) are in purple. The dashed curves are velocity nullclines, where $\ddot{x} = 0$, which provides a natural initial condition at any $x$.
  The critical initial conditions $x_c(t=0)$ leading to collisions are also shown (red points) for the Stokes flows in (a-b), and in (e); there is no such dependence on initial conditions in the inviscid case.
  }
  \label{fig:onedplot}
\end{figure*}

\subsection{Incompressible Stokes flow}

For Stokes flow impinging on a cylinder 
we have the Stokes flow field of \Eq{eq:xffs}, which when inserted into \Eq{eq:oned1} gives
\begin{equation}
     \St \, \ddot{x}+\dot{x} + x^2=0~~~
     \mbox{incompressible Stokes flow}
  \label{eq:oned1Stokes}
\end{equation}
We start by setting $\St = 1$. By following trajectories at differing initial conditions we can plot phase portraits in the $x-\dot{x}$ plane, see \partFig{fig:onedplot}{a}. Valle\'{e} \etal\ \cite{vallee2018} have obtained essentially the same phase portrait (the left panel of their Fig.~8).
There are two kinds of trajectories: The first kind (shown in yellow) asymptotically approach the marginally-stable fixed point at $x=\dot{x}=0$.
They therefore do not collide with the surface in finite time.
The second kind (purple) reach $x=0$ in finite time and with $\dot{x}<0$, and so collide with the surface. The two kinds of trajectories lie in attractor basins for, respectively, the fixed point at the origin (yellow), and a second attractor located at `$-\infty$' in $x$ and $u=\dot{x}$ (purple). This second attractor is inaccessible here due to collisions at $x=0$.

Now that we know there are two attractor basins, we ask how this phase-space geometry is affected by changing $\St$.
Increasing $\St$ in \partFig{fig:onedplot}{b} we see the noncolliding basin contracts as more initial collisions result in collisions.
This explains the threshold in inertia needed to see collisions: keeping the initial condition fixed, increasing $\St$ will transition the resulting trajectory from a noncolliding state to a colliding state when the basin shrinks away from the selected state.
This transition is discontinuous, but the time to collision diverges.

Despite the fact that we have plotted two phase portraits for the Stokes-flow case, there is in fact only one phase-portrait topology at all values of $\St$.  The Stokes number $\St$ can be eliminated from \Eq{eq:oned1Stokes} with the change of variables $x(t) = \widetilde{x}(\tau) / \St$ and $\tau = t / \St$ so that $\partial_t = \St^{-1} \partial_\tau$. Now we have
\begin{equation}\label{eq:perpendicular-eqn-rescaled}
  \partial_{\tau\tau}^2 \widetilde{x} + \partial_\tau \widetilde{x} + \widetilde{x}^2 = 0\,.
\end{equation}
This has yielded a universal equation for on-axis particles in incompressible stagnation point flows subject to a no-slip boundary condition.
Thus in the $\widetilde{x}$-$\widetilde{x}_\tau$ plane there is only one universal phase portrait, with one global topology. Changing the value of, for example $\St$, cannot change the phase portrait topology.
Note that in the generalised flow case the rescaling $x(t) = \widetilde{x}(\tau) / \St^{1/(m-1)}$ with $\tau$ above for $m \ne 1$ gives the same universal phase-portrait with respect to $\St$ for \Eq{eq:oned1}; the phase portraits shown in \partFig{fig:onedplot}{e}--\partfig{fig:onedplot}{f} for noninteger values of $m$ are also universal in this sense.

As the critical inertial value depends on the initial condition, the identification of a critical Stokes number depends on the relationship between $x(t=0)$ and $\dot{x}(t=0)$.
The most natural initial condition is for a particle to be dropped into the flow with zero initial acceleration $\ddot{x} = 0$, i.e.\ along the nullcline of the fluid flow velocity.
We have plotted the velocity nullcline as the dashed curve in \partFig{fig:onedplot}{a}--\partfig{fig:onedplot}{c}.
The critical value of $x(t=0; \St)$ then occurs where this nullcline intersects the colliding basin for the first time when moving out from the origin.
For $\St = 1$, we numerically find a critical initial condition of $x_c(t=0) \simeq 0.597$ by following trajectories backwards in time from the origin along the separatrix and observing where this intersects the nullcline.

Finally, we note that our rendered phase portraits were obtained using a small $x$ expansion.
We have performed similar calculations using a more complete flow field (the Kuwabara flow) which resulted in the same basic phase portraits and a threshold inertia.
As such, we believe the phase space geometry described above is universal for stagnation point flows at smooth obstacles with the no-slip boundary condition applied.

\subsection{Inviscid flow}

Here the stagnation point is very different to that in Stokes flow, the flow velocity varies linearly not quadratically with distance from the stagnation point \cite{turner2023,acheson_book}.
We then have the equation of simple harmonic motion (SHM), as many others have realised \cite{taylor1940,*taylor_vol3,langmuir1946,vallee2018}.
It is a classic physical problem to show that SHM permits two classes of solution: exponentially decaying solutions (overdamped) for $\St < \frac{1}{4}$ and periodic solutions (underdamped) for $\St > \frac{1}{4}$. These are in \partFig{fig:onedplot}{c}--\partfig{fig:onedplot}{d}.
The overdamped solutions only collide as $t\to\infty$ while the underdamped solutions collide in finite time.
So here the critical Stokes number $\Stcrit=1/4$ for a cylinder.
This was first realised by Taylor \cite{taylor1940,*taylor_vol3}, and by Langmuir and Blodgett \cite{langmuir1946}.
Note they actually quote $\Stcrit = 1/8$ which differs by a factor of 2 because of their alternative choice to set $U = U_\infty$ (see discussion at end of section~\ref{sec:model}).

At $\Stcrit$ there is a \emph{global} topology change in the phase portrait in the inviscid case, compare \partFig{fig:onedplot}{c} and \partfig{fig:onedplot}{d}. For $\St>\Stcrit$ all trajectories collide (are underdamped in SHM language).

\begin{figure}
  \centering
  \includegraphics[width=\linewidth]{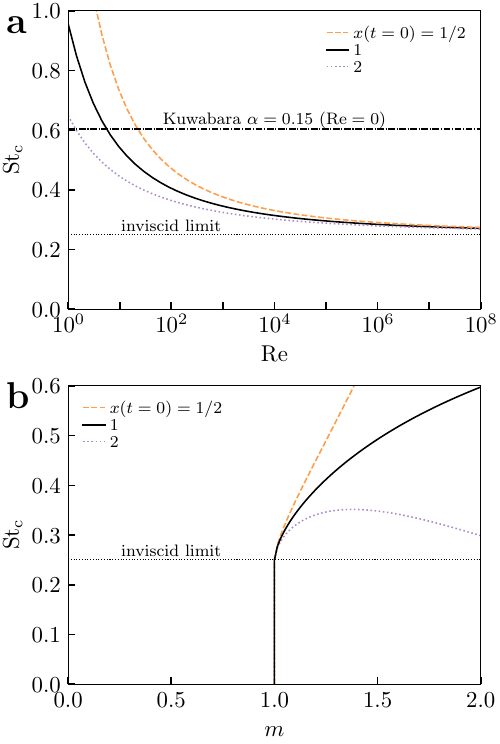}
  \caption{(colour online)
  Critical Stokes numbers above which collision occurs on-axis $y = 0$ for different initial conditions.
  (a) Stagnation point flows (Hiemenz flow) at finite Reynolds number $\Re$.
  (b) Power law flows $\vec{u} \propto (-|x|^m, m|x|^{m-1}y)^\top$, transitioning from Stokes to inviscid flows as $m$ decreases from 2 to 1.
  These both trend towards the inviscid limit $\Stcrit=1/4$ (dotted line) as $\Re \to \infty$ or $m \to 1$.
  }
  \label{fig:singular-threshold}
\end{figure}

\subsection{Different mechanisms for critical Stokes numbers in the Stokes-flow and inviscid-flow limits}
We can see from the phase portraits in \Fig{fig:onedplot} that there are two
completely different mechanisms for generating a critical Stokes number describing a dynamical transition.
For inviscid flow $\Stcrit=1/4$, a universal value is set by the stagnation-point flow alone, and $\Stcrit$ is set by a change in topology of the phase portrait, going from \partFig{fig:onedplot}{c} to \partfig{fig:onedplot}{d} \footnote{In the inviscid case, $\Stcrit=1/4$ is an upper bound.
If the initial speed $\dot{x}$ is made large enough, then a critical Stokes number dependent on initial conditions is obtained, just as in the Stokes-flow case.
However, our studies with global flow fields always found $\Stcrit=1/4$.}.
In this case, perturbing the flow field far from the stagnation point does not change the value of $\Stcrit$.
So in practice we expect that at high Reynolds number perturbing the global flow field far from the stagnation point should only very weakly alter the value of $\Stcrit$.

However for Stokes flow, the value of $\Stcrit$ depends on the initial conditions, and there is only one phase portrait topology (i.e.\ that of \partFig{fig:onedplot}{a}).
A consequence of this is that the numerical value of $\Stcrit$ will depend on the full flow field, not just the limiting flow at the stagnation point.
For example, perturbing the flow upstream by adding another cylinder would change the value of $\Stcrit$.

In \partFig{fig:singular-threshold}{b} we show results for the value of $\Stcrit$ as a function of the exponent $m$ in the distance dependence of the flow speed near the stagnation point.
We see that only for $m=1$ is there a universal value of $\Stcrit$. For larger values, $m>1$, there is a critical value of the Stokes number, but it will depend on the flow field far from the stagnation point.  So the value $m=1$ is the special case, as in \Fig{fig:onedplot}. When $m<1$ there are collisions at all values of the Stokes number and so no critical value.

\subsection{Hiemenz flows for large but finite Reynolds numbers}

We have seen that the inviscid, infinite Reynolds number, case is a special case. To obtain some insight into the behaviour at large but finite $\Re$ we use the Hiemenz flow field \cite{hiemenz1911, schlichting2017}.  This flow field is a simple model for flow at large $\Re$ near a stagnation point on a plane (so we cannot consider the effects of curvature).
The Hiemenz flow \cite{hiemenz1911} models stagnation point flows \emph{exactly}, albeit only asymptotically close to the dividing flow.
This flow features two parts: An inner part, the boundary layer, where the flow is akin to Stokes flow satisfying the no-slip boundary condition, and an outer part, where it is close to inviscid flow \cite{acheson_book}.
The boundary layer has a thickness that scales as $\Re^{-1/2}$, and so this thickness tends to zero as the Reynolds number increases.
See SM for details of this flow field.

We show the critical Stokes number $\Stcrit$ as a function of Reynolds number $\Re$ in \partFig{fig:singular-threshold}{a}.
Just as in the power law case, we recover the universal inviscid limit, and we see that as we move away from this, the critical Stokes number depends on the initial condition. Thus we expect that $\Stcrit$ is only set by the stagnation point flow in the limit of inviscid flow, but that at large Reynolds number, the dependence of the value of $\Stcrit$ on the flow field far from the stagnation point becomes weak.

As an example of Stokes flow, we also show $\Stcrit$
for the Kuwabara approximate flow of one cylinder in an array of cylinders. We studied this system in an earlier work \cite{robinson2021}. The value of $\Stcrit$ depends on the density $\alpha$ of the array.

\begin{figure}[b]
  \centering
  \includegraphics[width=\linewidth]{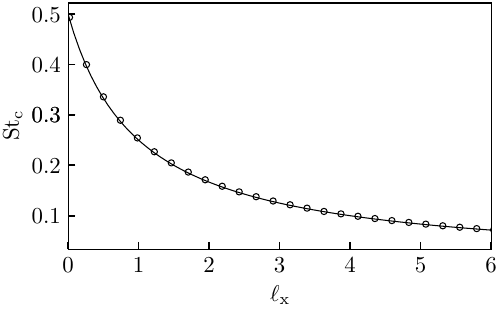}
  \caption{(colour online) The critical Stokes number for an ellipse in inviscid flow, as a function of the length of the ellipse along the flow direction. The length of the ellipse perpendicular to the flow direction is kept constant and is the length scale used to define the Stokes number. The points are the results of numerical calculations, and the curve is \Eq{eqn:ellipse_stc}. The initial conditions taken were $r(0)=10$, $\theta(0)=\pi$, $\dot{r}=u_r$ and $\dot{\theta}=0$ and full flow field (not \Eq{eq:xffs}) is used.
  }
  \label{fig:ellipse_Stc}
\end{figure}

\subsection{Ellipses in inviscid flow}
Finally we consider ellipses in inviscid flow, see \Fig{fig:ellipse}. The flow field near the stagnation point simply changes from $-x$ for a cylinder to $-\frac{1}{2}(1+\ell_x)x$ for an ellipse (see section 3 of the SM). For inviscid flow the flow near the stagnation point sets the value of $\Stcrit$. So for an ellipse in inviscid flow the critical Stokes number is simply
\begin{equation}
    \Stcrit=\frac{1}{2(1+\ell_x)}\,.
    \label{eqn:ellipse_stc}
\end{equation}
When $\ell_x=1$ we recover $\Stcrit=1/4$ for a cylinder.
The $\ell_x=0$ limit of the ellipse is a thin chord (wall) perpendicular to the flow, of length 2 in our non-dimensional units, and here $\Stcrit=1/2$, twice that for a cylinder. In the large $\ell_x$ limit the ellipse becomes very elongated along the flow direction, and the critical Stokes number decreases as $1/\ell_x$.
\Eq{eqn:ellipse_stc} is compared to numerical results in \Fig{fig:ellipse_Stc}. The agreement is excellent.

\section{Inertia-driven collisions off axis}
\label{sec:off-axis}
We now turn to particles approaching off axis ($\theta \ne \pi$). Here we are interested in the width of the collection window $\lambda$, within which particles collide with the obstacle and outside which they circumvent it. The width $\lambda$ is illustrated in \partFig{fig:kuwabara}{a}, as the distance between the limiting (dashed red) trajectories that just collide.
This width gives an estimate of the collection efficiency.

For limiting off-axis flows about the forward stagnation point, it is convenient to introduce a second variable $y = \ell_x (\theta - \pi)$ quantifying the azimuthal distance from the stagnation point.
In this section we focus on flows with circular symmetry $\ell_x = 1$ in the asymptotic limit approaching the forward stagnation point.
The particle equation of motion \Eq{eq:stokes-newton2} becomes
\begin{equation*}
\begin{split}
  \St \Bigl( \ddot{x} - (1 + x) \dot{y}^2\Bigr)
  &=
  - ( \dot{x} - u_r)\,,
  \\
  \St \Bigl( \ddot{y} + \frac{2}{1 + x} \dot{x} \dot{y}\Bigr)
  &=
  - \Bigl( \dot{y} - \frac{u_{\theta}}{1+x} \Bigr)\,.
\end{split}
\end{equation*}
Expanding around the stagnation point and discarding terms beyond leading order, we find this reduces to
\begin{equation}
\begin{split}
  \St ( \ddot{x} - \dot{y}^2)
  &=
  - ( \dot{x} - u_r )\,,
  \\
  \St ( \ddot{y} + 2 \dot{x} \dot{y})
  &=
  - ( \dot{y} - u_\theta )\,.
  \label{eq:xydot-asymptotic}
\end{split}
\end{equation}
approaching the stagnation point.
For the remainder of this section we will consider particle trajectories following \Eq{eq:xydot-asymptotic} using limiting flows we introduced in the previous section.
We will return to ellipses using the full (non-expanded) equation of motion \Eq{eq:stokes-newton2} in the next section.

For our chosen flows we determine the critical Stokes number $\Stcrit$ and the efficiency $\lambda$ numerically via an iterative procedure.
For a given initial condition, we assign a trial Stokes number $\St$ to the midpoint between the lower and upper bounds for $\Stcrit$ and follow the trajectory to test for a collision.
We update either the lower or upper bounds for $\Stcrit$ depending on whether collision occurs.
Each iteration therefore provides approximately 1 bit of information, which converges to sufficient precision for our plots after 20 iterations.
Once we have $\Stcrit$ we obtain $\lambda(\St)$ via a very similar procedure: for each $\St - \Stcrit$ we vary $y(t=0)$ to adjust the best upper/lower bounds for $\lambda$ at this value of $\St$.
We always assume the initial condition $x(t=0) = 1$ and $\dot{x}(t=0)=0$.

\subsection{Flow fields near stagnation points and the resulting equations for particle motion}

Here we use the Kuwabara flow field \cite{kuwabara1959} for flow around a cylinder in an array of cylinders representing e.g.\ porous media.
The area of this cylindrical pore is finite thus avoiding the Stokes paradox.
The full flow field is given by Eq.~(S1) in the SM.
Expanding the flow field around the stagnation point gives the limiting flow
\begin{equation}
  \vec{u} \propto (-x^2, 2 xy)^\top\,.~~~~
  \mbox{Stokes, cylinder}
\end{equation}
Due to the no-slip boundary conditions the flow speed tends to zero proportional to $x^2$.  The stagnation point is shown as a circle in \partFig{fig:kuwabara}{a}.

We now consider incompressible flow in two dimensions near a more general stagnation point. By `near' we mean that the distance to the stagnation point is much smaller than any other length scale in the system (e.g.\ the size of the obstacle or distance to another obstable).
In this case, when all other length scales can be discarded, we expect the flow field to have power law behaviour.
We write the resulting on-axis flow as $u_r\propto -x^m$, where the exponent $m\in[1,2]$ so as to provide a transition from the inviscid flow case ($m=1$) to the Stokes flow case $(m=2)$.

In two dimensions, the local incompressibility condition about the stagnation point in our $(x,y)$ coordinates amounts to \begin{equation}
    \partial_x u_r = -\partial_y u_\theta\,.
    \label{eq:incomp}
\end{equation}
The flows bifurcate at the stagnation point, so $u_\theta$ can only involve odd powers of $y$, with the leading term being linear $u_\theta \propto y$. But now that we have $u_r$, incompressibility \Eq{eq:incomp} gives us $u_\theta\propto - mx^{m-1} y$ with the same constant of proportionality as for $u_r$.
Combining the two components, the limiting flow field becomes
\begin{equation}
    \vec{u} \propto (-x^m, mx^{m-1}y)^\top ~~~ m>0~~~~
    \label{eq:general_stagff}
  \mbox{general}
\end{equation} 
which is therefore generally expected near the stagnation point, for incompressible two-dimensional stagnation point flows. Far from the stagnation point the flow field must relax to some background flow. 

Stagnation points of the general form of \Eq{eq:general_stagff} are found on a range of bodies in flowing fluids, with $m$ depending both on the geometry of the body and on the Reynolds number. For a cylinder we have $m=2$ for Stokes flow and $m=1$ for inviscid flow.
Flat walls belong to the same class as cylinders as they are obtained in the limit of infinite radius.
Similarly, any smooth convex body, such as an ellipse, belongs to this class as its local geometry will be indistinguishable from that of a cylinder with a finite radius of curvature. Thus, for example, we expect qualitatively the same behaviour for ellipses as for discs, but note that, as we can see from the factors of $\ell_x$ in Eq.~(S13) of the SM, details such as the value of critical Stokes number will change.

\subsection{Limiting power-law flows}
Here we present numerical data for the limiting power law flows $u = -|x|^m$ that we explored qualitatively (with phase portraits) in the previous section and in \Fig{fig:onedplot}.
Recall that $m = 2$ corresponds to a no-slip boundary condition (limiting Stokes flow), and $m = 1$ to a slip boundary condition (limiting inviscid flow), so non-integer values of $m$ interpolate these two limits.

We show how collection efficiency scales with $\delta = \St - \Stcrit$ in \partFig{fig:singular-onset}{b} for $m \in [1, 2]$.
We find that for small $\delta$ and $m > 1$, the collection efficiency obeys the relation
\begin{equation}\label{eq:regular-scaling}
  \lambda \propto \delta^{1/2}\,.
\end{equation}
This is exactly the exponent of one half that Ara\'{u}jo~\etal~\cite{araujo2006} estimated computationally for Stokes ($\mathrm{Re}=0$) flow. 
There is a simple argument for why we should expect $\lambda\propto\delta^{1/2}$.
This scaling is equivalent to $\delta \propto \lambda^2$, which is generic to any system with a Taylor series expansion for $\delta(\lambda)$ around $\lambda=0$, which has symmetry such that $\delta(\lambda)=\delta(-\lambda)$ (which eliminates the term linear in $\lambda$).
The cylinder and wedge are symmetric around $y=0$, so we have that symmetry.
In other words, due to the symmetry around the stagnation point, we only require that a Taylor series for $\delta(\lambda)$ exist.

The function $\delta(\lambda)$ describes the surface of intersection between the colliding manifold and the selected initial conditions; the Taylor series (and the scaling $\lambda \sim \delta^{1/2}$) breakdown when this surface becomes singular.
One important limit where this happens is towards inviscid flows as $\mathrm{Re} \to \infty$.
This inviscid case corresponds to $m = 1$ where slip occurs at the boundary for power law flows.
Vall{\'e}e \etal\ \cite{vallee2018} and Turner and Sear~\cite{turner2023} have found explicitly the leading order behaviour for inviscid flow is $\delta\propto 1/(\ln\lambda)^2$ (equivalent to $\lambda \sim \exp{\left(-1/\delta^{1/2}\right)}$)\,; this scaling is reproduced in \partFig{fig:singular-onset}{b} for $m=1$.
This analytic form does not have a Taylor series expansion around $\lambda=0$: the derivatives diverge.
So the Taylor series breaks down as $\Re\to\infty$, violating the assumption of our argument above.

The limiting power law flow $u = -|x|^m$ universally applies as we approach a stagnation point.
It is perhaps surprising that the entire phenomenology of a normal square root scaling \Eq{eq:regular-scaling} (and its breakdown in the inviscid limit) is captured by such a simple limiting flow. 
To confirm the universality of this phenomenology we will next turn to more realistic flows at finite Reynolds number.

\subsection{More realistic stagnation point flows}

In order to confirm that the limiting flows display the general phenomenology of realistic flows, we now consider idealised stagnation point flows.
In particular we consider the Stokes flow around a cylindrical collector in porous media (Kuwabara flow of \Refcite{kuwabara1959}) and the stagnation point flow onto a plane at non-zero Reynolds number $\Re$ (Hiemenz flow of \Refcite{hiemenz1911}).
The Hiemenz flow \cite{hiemenz1911} models stagnation point flows \emph{exactly}, albeit only asymptotically close to the dividing flow.
This flow features two parts: An inner part, the boundary layer, where the flow is akin to Stokes flow satisfying the no-slip boundary condition, and an outer part, where it is close to inviscid flow \cite{acheson_book}.
The boundary layer has a thickness that scales as $\Re^{-1/2}$, and so this thickness tends to zero as the Reynolds number increases.
See SM for details of this flow field.
For the cylindrical Kuwabara flow we integrate the polar equations given in \Eq{eq:stokes-newton2}, and for the Hiemenz flow we integrate the equivalent equation in Cartesian coordinates.

This means the previously-obtained critical value of $\Stcrit \simeq 0.597$ is not a universal feature of Stokes flows. The threshold for capture in Stokes flow depends (weakly) on details of the flow geometry.

\begin{figure}
  \centering
  \includegraphics[width=\linewidth]{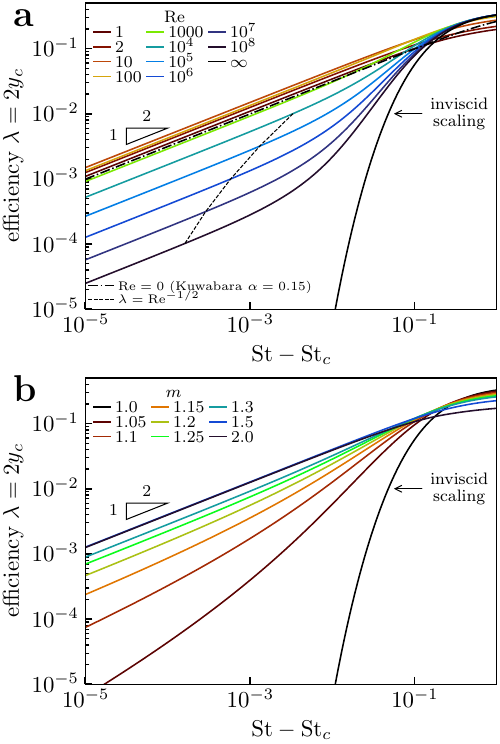}
  \caption{(colour online)
  Critical scaling of collection efficiency $\lambda$ above $\Stcrit$ with $x(t=0) = 1$.
  (a) Stagnation point flows at finite Reynolds number $\Re$ (Hiemenz flow).
  $\lambda$ shows square root scaling for efficiencies smaller than the viscous boundary layer thickness $\Re^{-1/2}$ (dashed line), and a cross-over to an inviscid-like scaling occurs at moderate Stokes numbers.
  This boundary layer vanishes as $\Re \to \infty$ rendering the critical scaling singular.
  For reference, we also show the results for the realistic Kuwabara flow field (dash-dotted lines) in the limit $\Re \to 0$ within porous media; this flow field seems to share similar properties to a Hiemenz flow with finite $\Re = \order{1}$.
  (b) Power law flows $\vec{u} \propto (-|x|^m, m|x|^{m-1}y)^\top$ captures the same phenomenology with a much simpler model.
  We observe square root scaling for small $\St - \Stcrit$ so long as $m > 1$, and we see a cross-over to an inviscid-like scaling at moderate Stokes numbers.
  }
  \label{fig:singular-onset}
\end{figure}

We show the efficiency of capture $\lambda(\delta)$ in \partFig{fig:singular-onset}{a}.
As before in the simplified flows, we see a region of $\delta$ values over which we see a $\delta^{1/2}$ scaling, and the width of this region decreases with increasing Reynolds number.
This is exactly what we would expect if a Taylor series expansion exists for particles in air flowing around a stagnation point in Stokes flow but {\em not} when the flow is inviscid.
We have therefore reproduced the same phenomenology for these realistic flows as we obtained for the simplified power law flows, with breakdown of the $\delta^{1/2}$ scaling only occurring in the inviscid limit.
This underscores the fact that the inviscid limit is truly singular, whereas other flows generically observe the critical efficiency scaling reported by Ara\'{u}jo~\etal~\cite{araujo2006}

\subsection{Ellipses in inviscid flow}
\label{sec:ellipse}

\begin{figure}
  \centering
  \includegraphics[width=\linewidth]{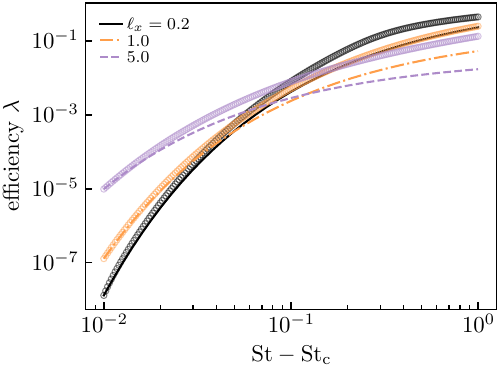}
  \caption{(colour online)
  Collection efficiency $\lambda$ as a function of Stokes number $\St$, for ellipses with different aspect ratios. The points are results of the numerical integrations, while the lines represent the analytical result in \Eq{eqn:ellipse_analytic}.}
  \label{fig:ellipse_lambda}
\end{figure}

Finally, we look at the effect of obstacle geometry on the collection efficiency $\lambda$. We do this by studying ellipses that are both elongated and flattened along the flow direction of inviscid flow.
The collection efficiency as a function of Stokes number is plotted in \Fig{fig:ellipse_lambda}. Results are plotted for both ellipses which are long along the flow direction, $\ell_x=5$, and for ellipses flattened along the flow direction, $\ell_x=0.2$.

As Vall{\'e}e \etal\ \cite{vallee2018} first showed for cylinders, for inviscid flow just above the 
$\Stcrit$, $\lambda$ has the unusual scaling $\lambda\propto\exp(-\delta^{-1/2})$. This is a much slower increase of $\lambda$ than the $\delta^{1/2}$ dependence found in Stokes flow. In \Fig{fig:ellipse_lambda} we see that the unusual scaling $\lambda\propto\exp(-\delta^{-1/2})$ also holds for ellipses in inviscid flow.

The precise theoretical behaviour of the collection  can be determined using the same analytical calculations laid out in Section V of Turner and Sear~\cite{turner2023} and we find that
\begin{equation}
\lambda\sim\exp\biggl(-\frac{(4-\sqrt{2})}{4\sqrt{1+\ell_x}}\pi\delta^{-1/2}\biggr),
\label{eqn:ellipse_analytic}
\end{equation}
The efficiency $\lambda$ is predicted to increase more rapidly with $\St$ for ellipses oriented with their minor axis along the flow direction (small $\ell_x$), than for ellipses oriented with their major axis along the flow direction.

From \Fig{fig:ellipse_Stc} we see that ellipses with small $\ell_x$ have large critical Stokes numbers, and from \Fig{fig:ellipse_lambda} we see that their collection efficiency increases rapidly above $\Stcrit$. So ellipses (in high Reynolds number flows) with small $\ell_x$ look good candidates for size-selective filters for particles. The sharper increase of $\lambda$ with $\St$, and hence particle size, means they are more selective than cylinders.

\section{Renormalisation-group derived series expansion results}
\label{sec:rg}

The equation of the damped harmonic oscillator \Eq{eq:stokes-newton-nondim} is readily solvable by undergraduates with some training in ordinary differential equations (ODEs).
By contrast, its deceptively simple extension to a quadratic restoring force (\Eq{eq:oned1} with $m = 2$) or for general $m$ requires a far more sophisticated treatment.
In approaching these problems notice that the limit $\St\to0$ is singular in the sense that $\St$ multiplies the highest-order derivative in \Eq{eq:stokes-newton-nondim};
the ODE is \emph{qualitatively} different in the case $\St \ne 0$ (where it is second-order) from $\St = 0$ (where it is first-order).
In addition, the time to reach the origin diverges as the critical trajectory is approached from the colliding manifold.
It turns out that this diverging timescale can only be captured by perturbation theories which treat the singular behaviour of $\St$.

A \naive perturbation theory constructed in powers of $\St\ll1$, which is only valid at short-times, therefore cannot capture the critical behaviour.
In particular, secular terms in perturbation theory cause the solution to diverge in the long-time limit.
Physically, this breakdown informs us that the relevant physics changes on different timescales.
This is formally identical to how inertia comes to dominate viscosity at large Reynolds number and thus fluid flow physics changes with length scale \cite{veysey2007}.
Viscosity is important within a `boundary layer' surrounding a surface, and unimportant far away.
In our case, we have a \emph{temporal} boundary layer: the flow field is unimportant within inertial timescales of $\order{\St^{-1}}$, and on longer timescales of $\order{1}$ particles `forget' their initial conditions and trace the flow field.
Asymptotic methods were specifically developed to resolve this kind of problem.

In a forthcoming publication \cite{robinsonRGFiltration2024}, two of us tease out the properties of the critical trajectory using the renormalisation group (RG) approach for solving singular differential equations \cite{chen1994,*chen1996,*kirkinisPRE2008}.
We are able to derive a series expansion for the critical trajectory of the on-axis problem of \Eq{eq:oned1} with $m=2$, and our end result is the following series for the critical value \cite{robinsonRGFiltration2024}
\begin{equation}
  (\St\, x)_\mathrm{c} 
  = 1 - \smallfrac{1}{2} + \smallfrac{1}{12} + \smallfrac{1}{48} - \smallfrac{1}{360} - \smallfrac{17}{4320} - \smallfrac{43}{80\,640} + \cdots\,.
\end{equation}
Numerical tests \cite{robinsonRGFiltration2024} indicate the series is convergent, and the limiting value $(\St\, x)_\mathrm{c}\simeq0.597777$ agrees with the result found in \partFig{fig:onedplot}{a}--\partfig{fig:onedplot}{b} by numerical integrating trajectories.
This work also demonstrates the usefulness of RG methods, which can readily handle the secular divergences at large times where standard perturbative methods fail.

\section{Conclusion}
We have studied an aerosol flowing over a simple model obstacle (a cylinder), and determined how and when the inertia of the aerosol particles causes these particles to deposit onto the obstacle.
For flow at all Reynolds numbers, there is a threshold amount of particle inertia below which no particles deposit.
However, we have shown (see \Fig{fig:onedplot}) that the dynamical transition that is behind this threshold is very different in the Stokes-flow and inviscid-flow limits.

In the inviscid limit, $\Stcrit$ is {\em only} a property of the stagnation point, while at any finite Reynolds numbers the threshold depends on the whole flow field, for example, in an array of cylinders, the threshold depends on the density of the array. Thus we expect the dependence of $\Stcrit$ on the flow field far from the stagnation point to be weak at large $\Re$ but significant for small $\Re$. 

Our calculations with Hiemenz flows suggests that $\Stcrit$ decreases with increasing $\Re$.
This matters for applications, as they span a huge range of Reynolds numbers from less than one for masks and other air filters \cite{robinson2021,riosdeanda2022}, to of order ten for beetles and small plants that harvest water from fog \cite{parker2001,*mitchell2020,*shahrokhian2020}, to thousands to millions for wind-turbine blades \cite{gao2021} and aircraft wings.

For inviscid flow, $\Stcrit$ is set by the stagnation point flow and we have analytic expressions for the flow field around an ellipse. we used them to show that stretching a cylinder along the flow direction to form an ellipse with its major axis along the flow direction reduces the critical Stokes number. Thus ellipses with long axes along the flow direction should be able to filter out smaller particles, or particles at lower air speeds, at least if the flow Reynolds number is significant. Drag at high Reynolds is approximately proportional to the area normal to the flow \cite{clancy1975}, so elongating the ellipse along the flow direction increases the size range of particles filtered, without increasing the drag. So if an application requires maximum filtration with minimum drag, elliptical obstacles are superior to cylinders.

\begin{acknowledgments}
\emph{Acknowledgements.}---JFR and RPS wish to thank Paddy Royall for igniting our original interest in the complex problem of particle capture during the early stages of the COVID-19 pandemic, and for many fruitful discussions.
JFR acknowledges funding from the Alexander von Humboldt foundation. MRT acknowledges funding from the EPSRC via grants EP/W006545/1 and UKRI070.
\end{acknowledgments}

\bibliography{selected}

\end{document}


\title{Supplemental Material}

\subtitle{Critical inertia for particle capture is determined by surface geometry at forward stagnation point}

\author{\small Joshua F. Robinson, Patrick B. Warren,
  Matthew R. Turner, and Richard P. Sear}

\date{\small\today}

\maketitle

\tableofcontents

\section{Stagnation point flows at cylindrical surfaces}
\label{sec:cylindrical-flows}

Below we derive the two `universal' stagnation point flow equations analysed in the main text for Stokes and inviscid flows, as limits of two well-known flows around cylinders.
While we derive these from cylindrical flows as convenient examples, the final stagnation point flows are expected to be independent of geometric details: the limiting flows depend on only (i) incompressibility and (ii) whether either a slip (inviscid) or no-slip (Stokes) boundary condition is applied.

\subsection{Stokes case ($\Re \to 0$): Kuwabara flow around a cylindrical collector}
\label{app:kuwa}
As is well known, there is no solution for Stokes flow around a single cylinder because the requisite far-field boundary conditions cannot be satisfied \cite{vandyke1975}, however a solution is possible in a system of \emph{many} parallel cylinders where the flow is more heavily constrained.
In the Kuwabara flow approximation the effective neighbourhood of the other cylinders is replaced by a zero-vorticity condition on an outer circular boundary $b$ around a single cylinder of radius $R < b$ \cite{kuwabara1959}.
This leads to a well-defined Stokes flow problem, and the approach has been widely used to model filtration problems with fibrous filters.

The resulting flow field in plane polar coordinates $(r, \theta)$ is given by 
\begin{subequations}\label{eq:kuwabara}
\begin{equation}
  \uhat_r = \frac{1}{r}\frac{\partial\psi}{\partial\theta}\,,\quad
  \uhat_\theta = -\frac{\partial\psi}{\partial r}\,,
\end{equation}
in terms of the stream function
\begin{equation}
  \psi=\frac{\Ub}{K}\biggl(
  r \ln{\Bigl(\frac{r}{R}\Bigr)}
  -\frac{(r^2-R^2)(r^2-R^2+2b^2)}{4rb^2}
  \biggr) \sin\theta
  \,,
\end{equation}
where $\Ub$ is the flow speed at $r = b$ (as well as the volumetric flow rate for the medium as a whole) \cite{kuwabara1959}.
The hats in $\hat{\uvec} = (\uhat_r, \uhat_\theta)^\top$ indicate that these are dimensional quantities, as opposed to the nondimensional $\uvec = \hat{\uvec} / U$ formed from dividing through by a characteristic speed $U$ to be specified below.
These equations are valid in the coaxial region $r \in [R, b]$.
The constant $K$ is the so-called \emph{hydrodynamic factor}, expressed in terms of volume fraction $\alpha = R^2/b^2 \in [0, 1]$ as
\begin{equation}
  K = -\frac{1}{2}\log{\alpha}-\frac{3}{4}+\alpha-\frac{1}{4}\alpha^2\,.
\end{equation}
Note that this flow field is not defined in the limit $\alpha \to 0^+$ because of the Stokes paradox alluded to above \cite{vandyke1975}.
\end{subequations}

Since the flow is left-to-right, the forward stagnation point is at $r=R$ and $\theta=\pi$.
To characterise the flow field in the vicinity of this point we therefore introduce the stagnation point coordinates $\hat{\rvec} := (\xhat, \yhat)^\top = (r - R, R(\theta - \pi))^\top$ as the dimensional counterparts of the nondimensional $\rvec := (x, y)^\top = \hat{\rvec}/R$ used in the main text.
Note that $\yhat$ is an arc length along the surface rather than an angle in these coordinates (and $y$ is simply the angle); this choice ensures $\yhat$ is dimensional so it is consistent with the Cartesian coordinate scheme used for e.g.\ Hiemenz flow (\cf section \ref{sec:hiemenz}).
Approaching the stagnation point $\uhat_{\xhat} \to \uhat_r$ and $\uhat_{\yhat} \to \uhat_\theta$.
Expanding $\hat{\uvec}$ from \Eq{eq:kuwabara} about the origin in this new coordinate system, we find to leading order
\begin{equation}
  \frac{\uhat_x}{\Ub} = \frac{\uhat_r}{\Ub}=-\frac{1-\alpha}{K}\,\frac{\xhat^2}{R^2} + \order{\textrm{cubic terms}}\,,\quad
  \frac{\uhat_y}{\Ub} = \frac{\uhat_\theta}{\Ub}=2\,\frac{1-\alpha}{K}\,\frac{\xhat\yhat}{R^2} + \order{\textrm{cubic terms}}\,.\label{sm:eq:kuwa1}
\end{equation}
i.e.\ only nondimensional quantities $(x, y)$ appear on the right-hand side.
In the main text our calculations are for $\alpha=0.15$, then we have
\begin{equation}
  \frac{\uhat_x}{\Ub} \simeq -2.48\,x^2 + \order{\textrm{cubic terms}}\,,\quad
  \frac{\uhat_y}{\Ub} \simeq 4.96\,xy + \order{\textrm{cubic terms}}\,.
  \label{eq:Kuwa_stag}
\end{equation}

Equation~\eqref{sm:eq:kuwa1} demonstrates that the (dimensional) flow field can be characterised as $\hat{\uvec} = (-\hat{k}x^2,2\hat{k}xy)^\top$ in the vicinity of the forward stagnation point in a cylindrical geometry, as claimed in the main text, and identifies the parameter as $\hat{k}=\Ub(1-\alpha)/K$ for the Kuwabara flow field.
This limiting flow field is actually expected for arbitrary (non-cylindrical) geometries, which we argue at the end of this section.

In practice, we select a characteristic speed $U$ to use a nondimensional $k = \hat{k} / U$ so $\uvec := \hat{\uvec} / U = (-kx^2, 2kxy)^\top$.
Working with the full Kuwabara flow field (which contains many higher-order terms), a natural choice for nondimensionalising the velocities is to take $U = \Ub$ giving $k = (1-\alpha)/K$ approaching the stagnation point.
We took $U = \Ub$ (with $k \simeq 2.48$ at $\alpha = 0.15$) for the numerical results  shown in Fig. 1 of the main text.
However, when we focus entirely on the limiting stagnation point flow without higher-order terms, all flow fields at different values of $k$ can be mapped onto each other simply by rescaling time. We thus have the freedom to choose $U = U_b (1-\alpha) / K$ so that $k = 1$ and $\uvec = (-x^2, 2xy)^\top$ without loss of generality.

\subsection{Inviscid case ($\Re \to \infty$): potential flow around a cylinder}

The flow of an inviscid, incompressible fluid around an infinitely long cylinder of radius $R$, is described by a simple analytic expression for the flow field \cite{acheson_book}.
This is a simple model of high-Reynolds-number flow.
Far from the surface of the cylinder the flow is homogeneous with magnitude $U_\infty$.
In plane polar coordinates $(r, \theta)$, the resulting flow field $\hat{\vec{u}}$ is given by
\begin{equation}
\frac{\uhat_r}{\Uinfty} = \left(1-\frac{R^2}{r^2}\right)\cos(\theta)\,,\quad
\frac{\uhat_\theta}{\Uinfty}=-\left(1+\frac{R^2}{r^2}\right)\sin(\theta),
\label{eq:potential-flow}
\end{equation}
using the same notation of the previous section where a hat (\,$\hat{}$\,) indicates the quantity is dimensional.

Repeating the procedure of the previous section, expanding around the stagnation point in \Eq{eq:potential-flow} gives to quadratic order in $x,y$
\begin{equation}
  \frac{\uhat_x}{U_\infty} = \frac{\uhat_r}{U_\infty} = -2x + 3x^2
  \,,\quad
  \frac{\uhat_y}{U_\infty} = \frac{\uhat_\theta}{U_\infty} = 2y - 2xy
  \,,
  \label{eq:flowfieldxy}
\end{equation}
where we have already changed to nondimensional coordinates $(x,y)^\top = \hat{\rvec}/R$.
The limiting flow field for inviscid systems is therefore $\hat{\uvec} = (-\hat{k}x, \hat{k}y)^\top$ in the vicinity of the forward stagnation point, where $\hat{k} = 2U_\infty$ from \Eq{eq:flowfieldxy}.
It is worth noting that as we do not have no-slip boundary conditions here, $u_{\theta}$ is not zero on the cylinder surface, except at the stagnation point; $u_{r}$ is zero at the surface because there is no flow into the cylinder.

Similar to the Kuwabara case, $U = \Uinfty$ is the natural choice for nondimensionalising when working with the full flow field.
As before, when just considering the limiting flow field the value of; $\hat{k}$ corresponds to an arbitrary time rescaling and so we can set $U = 2U_\infty$ so that $k = \hat{k}/U = 1$ without loss of generality.
In this case, the limiting nondimensional flow field becomes $\uvec = (-x, y)^\top$ which we used in the main text.
Note also that the critical value of Stokes number depends on the choice of $U$, so in this work we report it as $\Stcrit = \frac{1}{4}$ as we take $k = 1$.
However, elsewhere in the literature workers take $k = 2$ (from the choice $U = \Uinfty$) in which case $\uvec = (-2x, 2y)^\top$ and $\Stcrit = {1}/{8}$.
More generally, the critical solution for the on-axis equation of motion \[\St \, \ddot{x} + \dot{x} + k x = 0\] occurs at $k \, \Stcrit = {1}/{4}$.

\section{Stagnation point flows at arbitrary $\Re$: Hiemenz flow}
\label{sec:hiemenz}

Hiemenz flow is an exact solution to the Navier-Stokes equation in two-dimensions in the vicinity of an infinite flat wall.
For smooth walls, curvature corrections vanish on small length scales as $x \to 0^+$.
The Hiemenz flow therefore becomes (asymptotically) exact in the vicinity of the forward stagnation point.

It has velocity components $\vec{u} = (-\sqrt{(\nu k) }\phi(\eta), ky \phi'(\eta))^\top$, where $\nu$ is the kinematic viscosity, $k$ (with units of inverse time) characterises the flow velocity at the edge of the boundary layer, $\eta = \sqrt{k/\nu} \,x$ is a scaled distance from the surface, and the function $\phi(\eta)$ solves $\phi''' + \phi \phi'' - (\phi')^2 + 1 = 0$ with $\phi(0) = \phi'(0) = 0$ and $\phi'(\infty) = 1$ as boundary conditions.
We can take $\delta = \sqrt{\nu/k}$ as a proxy for the boundary layer thickness, so that the scaled distance variable $\eta = x/\delta$, leads to $\vec{u} = (-k\phi\delta, k\phi'y)^\top$.
To match to potential flow around a cylinder, we should have $k = 2U_\infty/R$ where $U_\infty$ is the flow in far-field and $R$ is the cylinder radius.
Then $\delta/R = \sqrt{\nu/(2U_\infty R)} = 1 / \sqrt{2\Re}$ where $\Re = U_\infty R/\nu$ is the Reynolds number.
For $\eta \to \infty$ the function becomes $\phi(\eta) = \eta - \eta*$ to leading order since $\phi'(\eta) \to 1$.
The constant of integration here $\eta* \simeq 0.6479$ is the `displacement thickness' in units of $\delta$.
In this limit therefore, $\vec{u} \to (-kx, ky)^\top$ as expected (indeed, by construction) -- this is matched to the potential flow to determine $k$.

Conversely, for $\eta \to 0$ the function must be $\phi(\eta) = \beta\eta^2$ to leading order because of the boundary conditions $\phi(0) = \phi'(0) = 0$.
Here $2\beta = \phi''(0) \simeq 1.2326$ is found by a shooting method; see also Table 5.1 in Schlichting \cite{schlichting2017}.
Therefore, in this limit $\vec{u} \to (-\beta kx^2/\delta, 2\beta k xy/\delta)^\top$, again displaying the expected dependency.

There are geometric limitations to this calculation, since for $\delta \ge R$ (corresponding to $\Re\leq1/2$) the Hiemenz flow field is invalidated by curvature effects (the boundary layer is bigger than the cylinder itself) and one should switch to another solution such as Kuwabara, or one of the various low $\Re$ approximations for flow around a cylinder instead.

\section{An ellipse in inviscid flow.}

The inviscid flow past an ellipse with equation
\begin{equation}
    \frac{X^2}{\ell_x^2}+Y^2=1,
    \label{eqn:ellipse}
\end{equation}
in an incoming flow of dimensionless speed $k/2$ is considered, where here $(X,Y)$ are the usual Cartesian coordinates. Here the ellipse has been non-dimensionalised to have thickness 2 in the stream-normal direction and has length $2\ell_x=\hat{\ell_x}/\hat{\ell_y}$ in the streamwise direction as per the main text. The flow field around the ellipse is found using standard potential flow theory and conformal mappings, for example see \cite{acheson_book}, section 4.8 for more details. 

In the physical $z$-plane, where $z=X+\ri Y$, the complex potential for the flow $w(z)$ is given by
\begin{equation}
w(z)=\frac{k}{2}\left(Z(z)+\frac{(1+\ell_x)^2}{4Z(z)}\right),
\label{eqn:w}
\end{equation}
where
\begin{equation}
z=Z(z)-\frac{(1-\ell_x^2)}{4Z(z)},
\label{eqn:c_map}
\end{equation}
is a conformal mapping which maps a circle of radius $\frac{1}{2}(1+\ell_x)$ in the $Z$-plane to the ellipse, \Eq{eqn:ellipse}, in the $z$-plane.

In order to calculate the flow velocities in the physical $z$-plane we need the inverse map of \Eq{eqn:c_map} which we find to be
\begin{equation}
    Z(z)=\left\{\begin{array}{ll} \frac{1}{2}\left(z+\sqrt{z^2+1-\ell_x^2}\right) & ~~~~\mathrm{for}~~~~~\Real(z)>0,\\
    \frac{1}{2}\left(z-\sqrt{z^2+1-\ell_x^2}\right) & ~~~~\mathrm{for}~~~~~\Real(z)<0
    \end{array}\right. ,
\end{equation}
where $\Real(z)=X$ the real part of $z$. Thus now the Cartesian velocity components $(u_x,u_y)$ is the $z$-plane are found from the complex potential in \Eq{eqn:w} as
\begin{eqnarray}
u_x-\ri u_y&=&\frac{dw}{dZ}\cdot\frac{dZ}{dz}, \nonumber \\
&=&\left.\frac{dw}{dZ}\right/\frac{dz}{dZ},\nonumber\\
&=&\left.\frac{k}{2}\left(1-\frac{(1+\ell_x)^2}{4Z(z)^2}\right)\right/\left(1+\frac{(1-\ell_x^2)}{4Z(z)^2}\right),
\label{eqn:flow_vel}
\end{eqnarray}
with $Z(z)$ given in \Eq{eqn:c_map}. Thus $u_x$ and $u_y$ are the real part and minus the imaginary part of this expression respectively.

In order to use these velocity components in Eq. (3) from the main paper we need to convert them to polar coordinates system, hence 
\begin{eqnarray}
    u_r&=&u_x\cos\theta +u_y\sin\theta ,~~~\\
    u_\theta&=&-u_x\sin\theta +u_y\cos\theta.
    \label{eqn:cart_polar}
\end{eqnarray}

The form of the flow in the vicinity of the stagnation point at $(X,Y)=(-\ell_x,0)$ can now be found from \Eq{eqn:flow_vel} by inserting
\[
z=(\ell_x+x)e^{\ri(\pi+y)},
\]
and carefully expanding for small $x$ and $y$ of the same order. In this way we find to leading order that
\[
z\sim-\ell_x-x-\ri \ell_xy,
\]
so
\[
4Z^2\sim (1+\ell_x)^2+2(1+\ell_x)^2+2\ell_x\ri (1+\ell_x)^2 y.
\]
Therefore
\[
u_x-\ri u_y\sim\frac{k}{2}(1+\ell_x)x+\frac{k}{2}\ell_x(1+\ell_x)\ri y.
\]
Hence expanding \Eq{eqn:cart_polar} for $\theta=\pi+y$ gives
\begin{eqnarray*}
    u_r\sim -u_x= -\frac{k}{2}(1+\ell_x)x\\
    u_\theta\sim-u_y= \frac{k}{2}\ell_x(1+\ell_x)y.
\end{eqnarray*}

Therefore $k=1$ and $\ell_x=1$ corresponds to the inviscid flow past a cylinder we consider in section 1.2 of this supplementary material.

\section{Equations of particle motion}

\subsection{Newton's equation for a suspended particle}

For a particle suspended in flowing air we assume that the only force on the particle is the friction with the surrounding air, which is taken to be proportional to the difference between particle's velocity $\hat{\vec{v}}$ and the local flow velocity $\hat{\vec{u}}$. This force is taken to act on the particle's centre of mass. Then
Newton's equation for the particle motion 
is
\begin{equation}\label{eq:stokes-newton}
  m \frac{\dd \hat{\vec{v}}}{\dd \hat{t}} = - \xi(\hat{\vec{v}} - \hat{\vec{u}})
\end{equation}
for a particle of mass $m$ and with drag coefficient $\xi$.
Again, a hat (\,$\hat{}$\,) indicates a dimensional quantity.
We switch to nondimensional quantities via transformations $\vec{r} = \hat{r} / R$, $\vec{u} = \hat{\vec{u}} / U$ and $t = U \hat{t} / R$, giving
\begin{equation}\label{eq:stokes-newton-nondim}
  \St \, \dot{\vec{v}} + \vec{v} - \vec{u}
  \equiv
  \St \, \ddot{\vec{r}} + \dot{\vec{r}} - \vec{u} = 0,
\end{equation}
where $\St = m U / (R \xi)$ is the Stokes number giving the effective inertia, and the dot (\,$\dot{}$\,) indicates a derivative with respect to nondimensional time $t$.

\subsection{Equation for particle near the cylinder's forward stagnation point}

We are interested in particle trajectories that pass close to the stagnation point at the front of the cylinder.
This stagnation point is at $r=1$ and $\theta=\pi$.
So we will study behaviour near this stagnation point.

In cylindrical polar coordinates we must include inertial terms in the acceleration term of \Eq{eq:stokes-newton-nondim}.
Decomposing $\vec{u} = u_r\vec{e}_r + u_\theta \vec{e}_\theta$ we find
\begin{equation}\label{eq:stokes-newton2}
  \St ( \ddot{r} - r \dot\theta^2 )
  =
  - ( \dot{r} - u_r )\,,
  \quad
  \St ( r \ddot{\theta} + 2 \dot{r} \dot{\theta} )
  =
  - ( r \dot{\theta} - u_\theta ).
 \end{equation}
We characterise the distance from the stagnation point via $(x,y)^\top = (r - 1, \theta - \pi)^\top$, giving
\begin{equation}
  \St ( \ddot{x} - (1 + x) \dot{y}^2 )
  =
  - ( \dot{x} - u_r ),
  \quad
  \St \Bigl( \ddot{y} + \frac{2}{1 + x} \dot{x} \dot{y}\Bigr)
  =
  - \Bigl( \dot{y} - \frac{u_{\theta}}{1+x} \Bigr).
  \label{eq:xydot}
\end{equation}
In our numerics with the Kuwabara flow field we integrate these equations with the exact flow field given in \Eq{eq:kuwabara}.

When we consider simplified flows approaching the stagnation point we expand in some small parameter $\epsilon$ where $x,y = \order{\epsilon}$.
We have to retain terms at least to $\order{\epsilon^2}$ in order to keep the equations coupled.
At this order we have
\begin{equation}
  \St ( \ddot{x} - \dot{y}^2)
  =
  - ( \dot{x} - u_r ),
  \quad
  \St ( \ddot{y} + 2\dot{x} \dot{y})
  =
  - ( \dot{y} - (1 - x) u_{\theta} ).
  \label{eq:xydot-expanded}
\end{equation}
The flow velocity terms must also be expanded to the same order.
In section \ref{sec:cylindrical-flows} we found $\vec{u} = (-k x^2, 2k xy)^\top$ as the limiting Stokes flow (no-slip) and $\vec{u} = (-k x + \frac{3}{2}k x^2, k y - k xy)^\top$ for inviscid flow (slip).
Note that the choice of $k$ corresponds to some arbitrary choice of the characteristic speed $U$ entering into the definition of $\St$, and so we have the freedom to choose $k=1$.
The first case then gives
\begin{equation}
  \St ( \ddot{x} - \dot{y}^2)
  =
  - ( \dot{x} + x^2 ),
  \quad
  \St ( \ddot{y} + 2\dot{x} \dot{y})
  =
  - ( \dot{y} - 2 xy ),
  \label{eq:xydot-stokes}
\end{equation}
and the inviscid case leads to
\begin{equation}
  \St ( \ddot{x} - \dot{y}^2)
  =
  - \Bigl( \dot{x} + x - \frac{3}{2} x^2 \Bigr),
  \quad
  \St ( \ddot{y} + 2\dot{x} \dot{y})
  =
  - ( \dot{y} - y + xy ),
  \label{eq:xydot-inviscid}
\end{equation}
at $\order{\epsilon^2}$.
Strictly speaking we must retain the $\order{\epsilon^2}$ flow-field terms on the right-hand side of \Eq{eq:xydot-inviscid}, but this makes no difference to the singular scaling above $\Stcrit$, and so we discard them in our numerics.
We must retain the $\order{\epsilon^2}$ inertial terms on the left-hand side so the equations remain coupled.

\section{Numerical calculations}

Particles are released into the flow  with zero initial acceleration ($\dot{\rvec}=\vvec=\uvec$).
The time-integration is performed using an RK45 method via the python implementation in \texttt{scipy.integrate.solve\_ivp}.

The critical Stokes number is obtained by setting $y = 0$.
Critical on-axis trajectories are determined by stepping off the origin in the direction of the unstable eigenvector and then integrating backwards in time.
The eigenvalue problem at the origin is nonlinear because of the quadratic vanishing of the flow field in the Stokes case.
The eigenvectors can be determined either by (i) taking the limit $x \to 0^+$ of a linear stability analysis at $x > 0$ or (ii) considering the solution to $\ddot{x} + \dot{x} = 0$ i.e.\ $(x, \dot{x})^\top = (A e^{-t}, -A e^{-t})^\top$.
By either route, the line $u = -x$ (for $\St = 1$) is obtained as the unstable eigenvector and must therefore be tangent to the critical trajectory approaching the origin.
We therefore begin the reverse-time integration beginning from some small perturbation $\vec{r} = (\epsilon, -\epsilon)^\top$ to obtain the critical trajectory for $\St = 1$, and thus the critical value of $x(t=0)$.
The critical trajectory at arbitrary $\St$ is then obtained by a trivial rescaling using the argument in section IV B of the main paper: this gives $\Stcrit$ once $x(t=0)$ is specified.

To calculate the efficiency we must integrate the full two-dimensional equations with $y \ne 0$.
In the $\St\to\infty$ limit, particles move in straight lines.
In that limit a cylinder of radius $R$ sweeps out a strip of air of thickness $2R$, collecting all the particles in this strip.
This allows us to define a deposition or collection efficiency $\eta = \lambda / R$ where $\lambda$ is the maximum initial orthogonal displacement (the $y$ coordinate in our notation) after which a particle would collide with the cylinder surface.
The displacement is taken far upstream of the cylinder.
The efficiency then corresponds to the fraction of particle trajectories within the column $R$ which would collide; it varies from zero when no particles collide, to one when $\St\to\infty$.

The fraction $\lambda / R$ does not easily generalise to other geometries, e.g.\ in the Kuwabara flow there is no obvious normalisation to replace $R$ in the confined pore geometry.
However, $\lambda$ can be generalised to arbitrary geometries as the measure of the number of streamlines (which has units of length in two dimensions).
Calculation of $\lambda$ is done by starting particle trajectories at varying values of the displacement normal to the flow direction.
We use an iterative technique to determine an upper and lower bound for $\lambda$, by varying the initial value of $y$ inside the best estimate.
If a particular choice of $y$ leads to a collision (non-collision), we increase (decrease) the respective bound.
We terminate the iterations once the difference between these bounds becomes sufficiently small.

\bibliographystyle{abbrv}
\bibliography{selected}